\documentclass[11pt]{article}

\usepackage{fullpage}
\usepackage{eucal}
\usepackage{algorithm2e}
\usepackage{times,amsmath,amssymb,xspace}
\usepackage{txfonts}
\usepackage{graphicx}
\usepackage{multirow}
\usepackage{amsmath}
\usepackage{amssymb}
\usepackage{theorem}
\usepackage{epsfig}
\usepackage{fancyvrb}
\usepackage{url}
\usepackage{cite}
\usepackage{multirow}
\usepackage{wrapfig}
\usepackage{comment}

\numberwithin{equation}{section}	% add the section number to the equation label

\theoremstyle{plain}			% use "default" font

\theoremstyle{definition}		% use "definition-style" font for the rest

\begin{document}
\title{STP/HAMPI for Computer Security}
\author{Vijay Ganesh\\
  {vganesh}@csail.mit.edu}

\maketitle

\abstract{In the past several years I have written two SMT solvers
  called STP and HAMPI that have found widespread use in computer
  security research by leading groups in academia, industry and the
  government. In this note I summarize the features of STP/HAMPI that
  make them particularly suited for computer security research, and a
  brief description of some of the more important projects that use
  them.}

\section{Introduction}

SMT solvers~\cite{smtlib, HandbookOfSAT2009}
(Satisfiability-Modulo-Theories Solvers) are computer programs that
decide the satisfiability problem for rich logics such as the theory
of bit-vectors and arrays~\cite{GaneshD07}, integers, and
datatypes. SMT solvers have recently proven to be particularly useful
in finding security vulnerabilities, debugging, and program analysis
aimed at security. The reason for the success of SMT solvers are
threefold: 1) The input logic of SMT solvers is rich enough to capture
a wide variety of program behavior easily and compactly, 2) SMT
solvers have become very efficient at solving such formulas obtained
from real-world applications, and 3) there are very effective
techniques now available, such as symbolic execution~\cite{exe, klee,
  dart}, that convert computation into SMT formulas. My solvers,
STP~\cite{GaneshD07} and HAMPI~\cite{hampi}, are specifically designed
to support computer security applications that perform security
analysis aimed at finding security vulnerabilities~\cite{prateek},
detecting malware~\cite{song} and constructing
exploits~\cite{brumley, avgerinos:2011}.

\section{STP}

STP~\cite{GaneshD07} is a solver for a theory of bit-vectors and
arrays. STP's logic is tailored to capture programs expressions
exactly. All modern computer program expressions can be reduced to
arithmetic and logic operations over suitably-sized (32 or 64 bit)
bit-vectors or read/write operations over memory. STP's logic of
bit-vectors captures program expressions, and STP's logic of arrays
captures memory read/writes. This exact bit-precision allows users to
easily encode a variety of security errors (e.g., off-by-one errors,
memory errors, overflow errors).

STP has been used in more than 100 research projects, a good number of
them are tools that automatically find security errors or perform
binary analysis. Important examples include: BitBlaze
project~\cite{song} from Dawn Song's group at Berkeley, The BAP system
from David Brumley's group at CMU~\cite{brumley:2011}, EXE~\cite{exe}
and KLEE~\cite{klee} from Dawson Engler's group at Stanford
University, S2E project~\cite{s2e} from George Candea's group at EPFL,
Switzerland, Akamai Inc. for finding security errors in mission
critical applications (contact: Michael Stone), and governmental
agencies. A comprehensive list of projects using STP can be found at
the following website:

\noindent{\url{http://sites.google.com/site/stpfastprover/tools-using-stp}}

\section{HAMPI}

HAMPI~\cite{hampi} is a solver for a theory of strings that can solve
constraints built out of string constants, variables, concatenation,
extraction and membership in regular expressions and context-free
grammars. HAMPI is explicity aimed at finding security
vulnerabilities, such as XSS attacks and SQL vulnerabilities, in web
applications written in JavaScript, PHP and Python.

The big users of HAMPI include: The Ardilla tool~\cite{symexe-icse09}
from Michael Ernst's group at MIT and University of Washington
Seattle, The WebBlaze project~\cite{prateek} from Dawn Song's group at
Berkeley, and Frank Tip's group~\cite{ArtziKDTDPE2010} at IBM
T.J. Watson center at Hawthorne in New York.

A comprehensive list of all the tools using STP and HAMPI can be found
by typing my name and following links at the Google Scholar's page:
\url{http://scholar.google.com}.

\bibliographystyle{abbrv}
\bibliography{stphampi-security}
\end{document}